\documentclass[a4paper,12pt,onecolumn,accepted=2021-06-23]{quantumarticle}
\pdfoutput=1

\usepackage{graphicx}
\usepackage[numbers,sort&compress]{natbib}
\usepackage[utf8]{inputenc}
\usepackage[english]{babel}
\usepackage[T1]{fontenc}
\usepackage{amsmath}
\usepackage{hyperref}
\usepackage{dsfont}

\newcommand{\bra}[1]{\ensuremath{\left\langle#1\right|}}
\newcommand{\ket}[1]{\ensuremath{\left|#1\right\rangle}}
\newcommand{\braket}[2]{\ensuremath{\left\langle#1 \vphantom{#2}\right| \left. #2 \vphantom{#1}\right\rangle}}
\newcommand{\bracket}[3]{\ensuremath{\left\langle#1 \vphantom{#2}\vphantom{#3}\right| #2 \vphantom{#1}\vphantom{#3} \left| #3  \vphantom{#1} \vphantom{#2}\right\rangle}}

\begin{document}

\title{Adiabatic critical quantum metrology cannot reach the Heisenberg limit even when shortcuts to adiabaticity are applied}
\author{Karol Gietka}
\orcid{0000-0001-7700-3208}
\email{karol.gietka@oist.jp}
\author{Friederike Metz}
\orcid{0000-0002-4745-7329}
\author{Tim Keller}
\orcid{0000-0003-4372-575X}
\author{Jing Li}
\orcid{0000-0002-7565-3933}
\address{Quantum Systems Unit, Okinawa Institute of Science and Technology Graduate University, Onna, Okinawa 904-0495, Japan}

\maketitle
\begin{abstract}
We show that the quantum Fisher information attained in an adiabatic approach to critical quantum metrology cannot lead to the Heisenberg limit of precision and therefore \emph{regular} quantum metrology under optimal settings is always superior. Furthermore, we argue that even though shortcuts to adiabaticity can arbitrarily decrease the time of preparing critical ground states, they cannot be used to achieve or overcome the Heisenberg limit for quantum parameter estimation in adiabatic critical quantum metrology. As case studies, we explore the application of counter-diabatic driving to the Landau-Zener model and the quantum Rabi model.
\end{abstract}

\section{Introduction}
Quantum metrology~\cite{giovannetti2006quantummetrology} is concerned with harnessing quantum resources following from the quantum-mechanical framework (in particular, quantum entanglement) to increase the sensitivity of unknown parameter estimation beyond the standard quantum limit~\cite{giovannetti2004quantum}. This limitation is a consequence of the central limit theorem which states that if $N$ independent particles (such as photons or atoms) are used in the process of estimation of an unknown parameter $\theta$, the error on average decreases as $\mathrm{Var}[\theta]\sim 1/\sqrt{N}$. Taking advantage of quantum resources can further decrease the average error leading to the Heisenberg scaling of precision $\mathrm{Var}[\theta]\sim 1/N$~\cite{giovannetti2004quantumenhanced} which in the optimal case becomes the Heisenberg limit $\mathrm{Var}[\theta] = 1/N$. This elusive limit~\cite{demkowicz2012elusive} has been a subject of intense experimental and theoretical research effort in the recent 20 years with both optical~\cite{DEMKOWICZDOBRZANSKI2015345,Polino2020photonicqm} and atomic~\cite{smerzi2018reviewatomquantum} systems. Unfortunately, highly non-classical states are also highly sensitive to decoherence and noise~\cite{escher2011generalnoisyqm,chaves2013noisyqm} which often renders quantum-enhanced measurements as proof-of-principle experiments. To counteract these effects, one can resort to quantum enhanced measurements without entanglement~\cite{braun2018withouthent} or critical quantum metrology associated with quantum phase transitions which received much attention in the recent years~\cite{paris2008spincriticality,paris2008qcriticalityresource,porras2013adiabaticqm,tsang2013qted,paris2014qmLMG,macieszczak2016dpt,paris2016dickqpt,porras2017quantumsensingcriticality,chitra2019qtransducerddpt,felicetti2020criticalqm,chu2021dynamicframeworkcqm}.

Critical quantum metrology~\cite{Zanardi2008QuantumCriticality} relies on the extreme sensitivity of equilibrium states near a critical point to small changes of physical parameters. In the vicinity of a critical point, such as a quantum phase transition, the ground state susceptibility can diverge leading to a possibility of arbitrarily precise estimation of certain physical parameters or, in other words, the super-Heisenberg scaling of the sensitivity~\cite{Rams2018atthelimitsofcriticality}, i.e., $\mathrm{Var}[\theta]\sim 1/N^x$ with $x>1$ or overcoming the Heisenberg limit $\mathrm{Var}[\theta]< 1/N$. However, if one takes into account the time needed to prepare such critical states, it turns out that the required time also diverges which is referred to as critical slowing down. As a result, the sensitivity of estimation in the critical case should be still bounded by the Heisenberg limit~\cite{Rams2018atthelimitsofcriticality}. A natural question that arises is whether one can use shortcuts to adiabaticity to arbitrarily reduce the time of preparing a critical state and eventually overcome the Heisenberg limit. Since realizing finite time adiabatic dynamics comes with an energy cost~\cite{abah2019energetic,muga2019shortcutsreview}, energy can then be treated as a resource in quantum metrology. This can be phenomenologically understood if one looks at the time-energy uncertainty principle $\mathrm{Var}[t]  \geq \hbar / \mathrm{Var}[E]$, where $\mathrm{Var}[t]$ is interpreted as the time a quantum system needs to evolve from an initial to a final state~\cite{busch2002time}.

Shortcuts to adiabaticity~\cite{muga2019shortcutsreview} is a framework devoted to finding fast routes to the final results of adiabatic changes of the system control parameters, and encompasses a variety of techniques including invariant based inverse engineering~\cite{lewis1982direct}, counter-diabatic driving~\cite{berry2009transitionless,demirplak2003adiabatic}, the fast-forward approach~\cite{masuda2010fast}, alternative shortcuts through unitary transformations~\cite{muga2010fastoptimalcoolingharmonic,torrontegui2011fastatomictransport,chen2010shortcut23}, and optimal control theory~\cite{torrontegui2013shortcuts}. From the viewpoint of metrology, counter-diabatic driving~\cite{Jarzynski2013,delcamp2013cddrive} seems enticing as it forces the state to be an instantaneous eigenstate of the bare Hamiltonian throughout the entire process by suppressing any transitions between the system’s eigenstates during the Hamiltonian dynamics.

In quantum metrology counter-diabatic-like techniques have already been applied to the estimation of parameters in time-dependent Hamiltonians by forcing the time-evolved state to maximize the quantum Fisher information at every point in time~\cite{pang2017optimaltd, cabedo2020shortcut,hou2021superandnotsuper}. However, here we consider the case where the part of the Hamiltonian that depends on the unknown parameter is time-independent. Moreover, we utilize counter-diabatic driving in its traditional sense that is, for ground state preparation. Specifically, our goal is to drive the system to the ground state near the critical point of the Hamiltonian starting from an uncritical state. The driving gives rise to an overall time-dependent Hamiltonian, but the mechanism imprinting the information about the unknown parameter remains time-independent.

In this work we show that critical quantum metrology cannot beat the Heisenberg limit if the unitary dynamics required for critical ground state preparation is taken into account which is often overlooked in studies of critically enhanced metrology protocols. We therefore argue that given the same amount of time resources \emph{regular} quantum metrology always gives rise to better sensitivities than critical quantum metrology. The latter in general diverges for critical ground state preparation due to the problem of critical slowing down. To avoid the problem of diverging state preparation times we consider shortcuts to adiabaticy, specifically counter-diabatic driving, and apply these tools to two systems studied in the context of criticality and metrology. The first system is the Landau-Zener model~\cite{yang2017LZmetrology,campbell2017shortcutstradoffLZ} which we treat as a toy model for criticality, and the second one is the quantum Rabi model~\cite{felicetti2020criticalqm,chen2021shortcutstaqrm}. Subsequently, by a careful analysis of the quantum Fisher information, we argue why counter-diabatic driving cannot be used to achieve the Heisenberg limit in critical quantum metrology and in general gives rise to a smaller quantum Fisher information than purely adiabatic state preparation.


\section{Preliminaries}\label{sec:1}
\subsection{Quantum metrology}\label{sec:qm}
Quantum metrology is based on the quantum theory of estimation~\cite{helstrom1976quantum} which states that when estimating an unknown parameter $\theta$, the sensitivity $\mathrm{Var}[\theta]$ is limited by the quantum Cram\'er-Rao bound~\cite{caves1994geomqs}
\begin{align}
    \mathrm{Var}[\theta] \geq \frac{1}{\sqrt{ \mathcal{I}_{\theta}}},
\end{align}
where $\mathcal{I}_{\theta}$ is the quantum Fisher information (QFI) which for pure states $|\psi\rangle$ is defined as~\cite{caves1994geomqs}
\begin{align} \label{eq:QFIgeneral}
    \mathcal{I}_{\theta} \equiv 4\left(\langle\partial_\theta \psi |\partial_\theta \psi\rangle - \langle \partial_\theta \psi| \psi\rangle^2 \right),
\end{align}
with $\partial_\theta \equiv \partial/\partial \theta$. If the unknown parameter is a quantum phase imprinted via a coherent dynamical process then $\theta = \omega T$ with $T$ being the total time of the process and $\omega$ being the unknown Hamiltonian parameter. In this case, the QFI gains time dependence ($\mathcal{I}_\omega = T^2 \mathcal{I}_\theta$) which means that by performing a long-enough measurement under idealistic conditions, one can obtain an arbitrary precision of estimation. Therefore, the time is considered a resource in quantum metrology. 

We now assume that the mechanism imprinting the information about an unknown parameter $\omega$ is known and can be described via a Hamiltonian of the form $\hat H = \omega \hat H_\omega + \hat H_t(t)$, where $\hat H_t(t)$ represents a general unknown-parameter-independent Hamiltonian while $\omega\hat H_\omega$ is the term imprinting the information about $\omega$ ($\hat H_\omega$ itself does not depend on $\omega$). Starting from an initial state $\ket{\psi_0}$, the time-evolved state after a time $T$ is given by (we set $\hbar =1$ throughout the entire manuscript)
\begin{align}\label{eq:psifinal}
 |\psi_f\rangle = \hat U|\psi_0\rangle = \mathcal{T}\exp\left(-i\int_0^T \hat H \mathrm{d}t\right)|\psi_0\rangle,
\end{align}
with $\mathcal T$ being the time ordering operator. This allows us to rewrite the expression for the QFI in the following way 
$\mathcal{I}_\omega \equiv 4\left(\langle \psi_0|\hat h^2|\psi_0\rangle - \langle\psi_0|\hat h|\psi_0\rangle^2 \right)$,
where $\hat h = i\hat U^\dagger\partial_\omega\hat U $. Moreover, it can be shown that the QFI is upper bounded by~\cite{caves2007generallimits} 
\begin{align} \label{eq:QFIlimit}
     \mathcal{I}_\omega  \equiv 4\left(\langle \psi_0|\hat h^2|\psi_0\rangle - \langle\psi_0|\hat h|\psi_0\rangle^2 \right)\leq 4T^2\left(\langle \psi_0|\hat H_\omega^2|\psi_0\rangle - \langle\psi_0|\hat H_\omega|\psi_0\rangle^2 \right).
\end{align}
For example, for two-mode systems \cite{pezze2018quantummetrologyreview} with a fixed total number of particles $N$, the right hand side of the above inequality can be maximized by using maximally entangled initial states which are in a superposition of eigenstates of the operator $\hat h$ with smallest and largest eigenvalue leading to the celebrated Heisenberg limit (HL) $\mathrm{Var}[\omega] = 1/NT$~\cite{caves1994geomqs}. However, for non-entangled initial states, the above inequality gives rise maximally to the standard quantum limit (SQL) $\mathrm{Var}[\theta] = 1/\sqrt{N}T$. Therefore, the quantum-enhanced sensitivity can be used as an entanglement witness~\cite{hyllus2012fiherentanglement,toth2012multihigh,strobel2014fisher}. For clarity, throughout this manuscript, we will refer to the above scenario as \emph{regular} quantum metrology and compare this scheme to the case of critical quantum metrology by carefully taking into account the time resources of either of the two approaches.

\subsection{Critical quantum metrology}\label{sec:cqm}
Critical quantum metrology takes advantage of criticality associated with continuous quantum phase transitions, for which in the thermodynamic limit the energy gap above the ground state closes~\cite{sachdev2007quantum}. The effect of a vanishing energy gap can be explicitly seen if we calculate the QFI for the ground state $|\psi_0(\omega)\rangle$ of the Hamiltonian $\hat H(\omega) = \sum_{n=0}E_n(\omega)|\psi_n(\omega)\rangle\langle\psi_n(\omega)|$~\cite{you2007fidelitycriticality}
\begin{equation}
    \mathcal{I}_\omega = 4 \sum_{n\neq 0} \frac{|\langle \psi_n(\omega) |\partial_\omega \hat H(\omega)| \psi_0(\omega)\rangle|^2}{[E_n(\omega)-E_0(\omega)]^2},
\end{equation}
which coincides with the real part of the quantum geometric tensor~\cite{zanardi2007qcstensor}. From the expression above, it is clear that if the energy gap above the ground state closes, the QFI diverges due to the vanishing denominator. This property may lead, in principle, to an arbitrarily high estimation precision in the thermodynamic limit. However, the critical unknown-parameter-dependent ground state has to be prepared beforehand. A standard approach to ground state preparation uses adiabatic time evolution in which an easy-to-prepare ground state is adiabatically evolved into the desired target ground state through a change in the Hamiltonian parameters. The adiabatic theorem states that a quantum system will remain in its ground state as long as it is changed slowly enough such that no excitations occur. However, this implies that near a quantum phase transition, where the energy gap vanishes, any change in the system parameters has to be infinitely slow---a behaviour known as critical slowing down. Therefore, a critical quantum metrology protocol which relies on preparing a ground state far away from the critical point and subsequently driving it close to the critical point~\cite{felicetti2020criticalqm} will inevitably require long preparation times which is often overlooked when calculating the achievable sensitivities and when comparing the approach to the \emph{regular} quantum metrology scheme. In fact, the QFI in critical quantum metrology is constrained to a tighter limit than the QFI in the \emph{regular} quantum metrology scheme, which we briefly outline in the following.

Note that adiabatic state preparation is still a unitary dynamical process and hence the final (critical) ground state is obtained via
\begin{align}\label{eq:finalstatecqm}
    \ket{\psi_f} \equiv \ket{\psi_f(\omega,g_f)} = \hat U(T,\omega,g_0,g_f) \ket{\psi_0},
\end{align}
where $\ket{\psi_0}\equiv\ket{\psi_0(\omega,g_0)}$ is the initial state that can in general depend on the unknown parameter $\omega$ and $g_0$, $g_f$ are the initial and final values of a control parameter $g$ which is adiabatically changed from an uncritical to a critical point of the Hamiltonian $\hat H = \omega \hat H_\omega + \hat H_t[g(t)]$. The QFI is calculated with respect to the final critical states
\begin{align}\label{eq:qfifinal}
    \mathcal{I}_{\omega} = 4\left( \braket{\partial_\omega \psi_f}{\partial_\omega \psi_f} -\braket{\partial_\omega \psi_f}{\psi_f}^2\right).
\end{align}
If we substitute equation \eqref{eq:finalstatecqm} into equation \eqref{eq:qfifinal} it is straightforward to show that the expression for the QFI consists of three terms (see Appendix \ref{appendix:A})
\begin{align}\label{eq:qfi3parts}
    \mathcal{I}_{\omega} = \mathcal{I}_{\omega}(\partial_\omega \hat U) + \mathcal{I}_{\omega}( \ket{\partial_\omega \psi_0}) + \mathcal{I}_{\omega}(\partial_\omega \hat U, \ket{\partial_\omega \psi_0}),
\end{align}
where the first term depends only on $\partial_\omega \hat U$, the second term depends only on $\ket{\partial_\omega \psi_0}$ and the third term depends both on $\partial_\omega \hat U$ and $\ket{\partial_\omega \psi_0}$. However, if the initial state is not critical itself, as in the protocols studied in this work, the dependence of the initial ground state on the unknown parameter can be neglected and the QFI becomes
\begin{align}
    \mathcal{I}_\omega = 4\left(\langle \psi_0|\hat h^2|\psi_0\rangle - \langle\psi_0|\hat h|\psi_0\rangle^2 \right),
\end{align}
with $\hat h = i\hat U^\dagger\partial_\omega\hat U$ which coincides with the expression for the QFI in equation \eqref{eq:QFIlimit} and is therefore upper bounded by the same limits, i.e.~the SQL for non-entangled states and the HL for maximally entangled states. However, as was previously shown, saturating the upper bound in equation \eqref{eq:QFIlimit} requires different states than the instantaneous ground states of the Hamiltonian $\hat H = \omega \hat H_\omega + \hat H_t[g(t)]$~\cite{pang2017optimaltd} and therefore critical quantum metrology is always inferior to the optimal \emph{regular} quantum metrology scheme when adiabatic ground state preparation is taken into account as well. Hence, the maximal attainable QFI $\mathcal{I}_\omega^c$ for a critical metrology protocol, i.e., when the time evolved state is the instantaneous ground state of the Hamiltonian $\hat H = \omega\hat H_\omega + \hat H_t[g(t)]$ obeys
\begin{align}\label{eq:crit}
	\mathcal{I}^c_\omega  < \max_{\{|\psi_0\rangle\}} 4\tau_c^2\left(\langle \psi_0|\hat H^2_\omega|\psi_0\rangle - \langle\psi_0|\hat H_\omega|\psi_0\rangle^2 \right),
\end{align}
where $\tau_c$ is the duration of the adiabatic protocol.

\subsection{Shortcuts to adiabaticity}
Shortcuts to an effective adiabatic dynamics can be realized with several different techniques~\cite{torrontegui2013shortcuts}. Here, we use counter-diabatic (CD) quantum driving as it ensures transition-less dynamics~\cite{delcamp2013cddrive}. This technique relies on adding an extra term to the original Hamiltonian $\hat H$ which guarantees time evolution corresponding to the adiabatic dynamics but within an arbitrary time. The CD term can be chosen as~\cite{santos2017generalized} 
\begin{align}
    \hat H_{\mathrm{CD}} = i \sum_n |\partial_t \psi_n(t)\rangle \langle \psi_n(t)|,
\end{align}
where $|\psi_n(t) \rangle$ are the instantaneous eigenstates of the original Hamiltonian. In general, the CD term might be non-local~\cite{muga2010transitionless,zurek2012criticalpoint} and thus often impractical for experimental realizations. Here, however, we are interested in the fundamental limitations under idealistic conditions, therefore a possible non-locality of the CD term does not constitute any issue.

Note that with the addition of the CD term the unitary evolution according to equation \eqref{eq:psifinal} is altered to 
\begin{align}
\begin{split}
	  |\psi_f(T, \omega, \tilde{\omega}, g_f)\rangle &=  \hat{U}_\mathrm{CD}(T,\omega, \tilde \omega, g_0,g_f)|\psi_0\rangle\\
	& =  \mathcal{T}\exp\left(-i\int_0^T \left(\omega\hat H_\omega + \hat H_t[g(t)] + \hat H_\mathrm{CD}[\tilde{\omega},g(t)]\right) \mathrm{d}t\right)|\psi_0\rangle .
	\end{split}
\end{align}
The appropriate CD term ensuring transitionless driving will in general depend on the parameters of the bare Hamiltonian including the unknown parameter $\omega$ either explicitly or implicitly through the chosen ramp function $g(t,\omega)$. However, the CD term constitutes a control term that has to be engineered and applied externally to the system of interest. Therefore, in any experimentally realistic setting the CD term can only depend on an initial estimate $\tilde \omega$ of the unknown parameter rather than on the unknown parameter $\omega$ itself. The estimate $\tilde \omega$ has to be updated after each measurement yielding better control terms and hence giving rise to an adaptive metrology scheme~\cite{pang2017optimaltd}. The fact that the CD term does not depend on the true unknown parameter but its estimate $\tilde \omega$ allows us to include the CD term into the general time dependent term $\hat H_t(t)$ of the Hamiltonian. Therefore, the arguments of the previous section still hold and the limitation from equation \eqref{eq:crit} is still valid. Thus, even if the estimate $\tilde\omega$ matches the true value of the unknown parameter to arbitrary precision, i.e.~$\tilde \omega = \omega$, and therefore the critical ground state is reached with perfect fidelity in an arbitrary short time using CD driving, the QFI is still bounded by the HL. Note that this statement holds for any form of applied optimal control in which the additional control term only depends on an estimate of the unknown parameter rather than on the unknown parameter itself. While these arguments already show that adiabatic critical quantum metrology with and without CD driving cannot beat or even reach the HL, they do not yet quantify the extent to how much better or worse one performs over the other. Therefore, we will explore two different examples in Section~\ref{sec:3} comparing the achieved QFI of the \emph{regular} and the critical quantum metrology approach with and without CD driving. 

\subsection{Fidelity and quantum speed limit}
In order to quantify whether a desired target (ground) state $| \psi_{t} \rangle$ is reached, we will use the fidelity defined as
\begin{align}
    \mathcal{F} = |\langle \psi_{f} | \psi_{t} \rangle|^2,
\end{align}
where $| \psi_{f} \rangle$ is the final state obtained after time evolution with the Hamiltonian $\hat H + \hat H_{\mathrm{CD}}$. We will also refer to the quantum speed limit (QSL) time~\cite{deffner2017quantumsl} which for a general Hamiltonian $\hat H$ is given by~\cite{PhysRevA.67.052109}
\begin{align}
  \tau_\mathrm{QSL} = \mathrm{max}\Bigg\{\frac{\hbar}{\mathrm{Var}[\hat H]} \arccos|\langle{\psi_0|\psi_t}\rangle|, \frac{2\hbar}{\pi \langle \hat H \rangle} \arccos|\langle{\psi_0|\psi_t}\rangle|^2 \Bigg\} 
\end{align}
and constitutes the fundamental maximum rate for quantum time  evolution.
%
%
\section{Case studies}\label{sec:3}
%
\subsection{Landau-Zener model}
\begin{figure}[htb!]
  \centering
\includegraphics[width=1\textwidth]{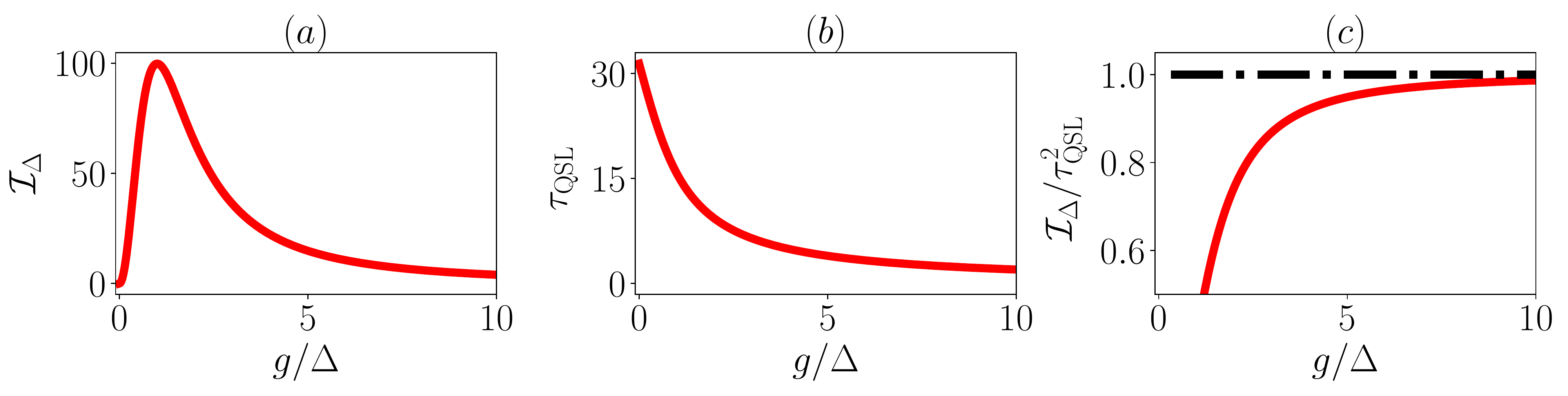}
\caption{Panels (a) and (b) present the QFI $\mathcal{I}_\Delta$ and the QSL time $\tau_{\mathrm{QSL}}$ for the LZ model as a function of $g/\Delta$, respectively. Panel (c) shows the QFI normalized to the square of the QSL time (solid red line) and the SQL (dash-dotted black line). In the simulations we set $\Delta = 0.05$.}
\label{fig:fig1}
\end{figure}

The Landau-Zener (LZ) model describes a two-level system in a time-dependent or controlled field $g(t)$~\cite{zener1932non}
\begin{align}\label{eq:LZ}
    \hat H_{\mathrm{LZ}} = \frac{\Delta}2 \hat \sigma_x + \frac{g(t)}2 \hat \sigma_z.
\end{align}
where $\Delta$ is the level splitting for $g = 0$, and $\hat \sigma_i$ is the $i$th Pauli matrix. Being a single-particle system, the LZ model does not exhibit a quantum phase transition, however, it features an avoided crossing at $g=0$ and can therefore be considered as a toy model for criticality~\cite{innocenti2020ultrafast,zhang2009directcriticality}. The instantaneous ground state of the LZ model is given by
\begin{align}
  |\mathrm{\psi_0}\rangle = \frac{g-\sqrt{\Delta^2+g^2}}{\sqrt{ 2g \left(g-\sqrt{\Delta^2+g^2}\right)+ 2\Delta^2}}|\!\downarrow\,\rangle+ \frac{\Delta}{\sqrt{ 2g \left(g-\sqrt{\Delta^2+g^2}\right)+ 2\Delta^2}} |\!\uparrow\,\rangle,
\end{align}
and the QFI with respect to an unknown parameter, which we set to $\Delta$, is calculated to
\begin{align}\label{eq:qfilz}
    \mathcal{I}_\Delta = \frac{g^2}{\left(\Delta^2+g^2\right)^2},
\end{align}
by using the definition from equation \eqref{eq:QFIgeneral}. We plot the dependence of the QFI on the control parameter $g$ in figure \ref{fig:fig1}(a), where we have set $\Delta=0.05$ for the unknown parameter. The QFI reaches its maximum value $\mathcal{I}_\Delta = 1/(4\Delta^2)$ close to the point of the avoided crossing at $g=\Delta$ and diverges for $(\Delta,g) \ll 1$. The divergent behavior of the QFI seems promising and suggests that arbitrary high sensitivities can be realized. However, achieving these high sensitivities requires the preparation of the critical ground state in the first place, which will inevitably suffer from the critical slowing down. Therefore, when comparing the critical quantum metrology scheme to the \emph{regular} scheme we also have to take into account the amount of time it takes to prepare the critical state. In what follows we assume that the initial state of the system is the ground state with $g \gg \Delta$, i.e.~the spin-down state $|\psi_0\rangle = |\!\downarrow \,\rangle$, and subsequently the control field $g(t)$ is adiabatically decreased to reach the ground state at $g = \Delta$. The required time for adiabatic state preparation with the LZ Hamiltonian of equation \eqref{eq:LZ} can be lower bounded by the QSL time which for the LZ can be calculated according to~\cite{bason2012high,hegerfeldt2013qsltime}
\begin{align}
    \cos\left(\frac{\Delta}{2}\tau_{\mathrm{QSL}}\right) = |\langle \psi_0|\psi_t(g) \rangle|,
\end{align}
where $|\psi_0\rangle$ and $|\psi_t \rangle$ is the initial and target ground state, respectively. Figure \ref{fig:fig1}(b) shows the QSL as a function of $g$ and illustrates the critical slowing down close to the avoided crossing.

The QFI for a single-particle system in the \emph{regular} quantum metrology setting under optimal conditions is given by the SQL (for the LZ model the SQL is also the HL since $N = N^2 = 1$) which for the LZ model takes the simple form $\mathcal{I}_\Delta^{\mathrm{SQL}} = T^2$ with $T$ being the total evolution time~\cite{yang2017LZmetrology}. Figure \ref{fig:fig1}(c) compares the QFI attained in the critical quantum metrology framework [equation \eqref{eq:qfilz}] to the SQL (black-dashed line) when using the same time resources, i.e.~$\mathcal{I}_\Delta^{\mathrm{SQL}} =\tau_{\mathrm{QSL}}(g)^2$. The performance of critical quantum metrology in comparison to the SQL becomes worse as $g$ is decreased towards the point of the avoided crossing. Hence, given the time it would take to prepare a critical ground state, one always achieves higher sensitivities by performing \emph{regular} quantum metrology (under ideal conditions) in the same amount of time.

Note that the QFI from equation \eqref{eq:qfilz} asymptotically reaches the SQL for $g\gg \Delta$. The SQL (the maximum QFI) is obtained when the evolved state is in a superposition of eigenstates with highest and lowest eigenvalue of the operator $\hat h = i\hat U^\dagger\partial_\Delta\hat U $ at all times~\cite{pang2017optimaltd}. In the limit $g\gg \Delta$, i.e.~far away from the avoided crossing, the required time for adiabatic state preparation and therefore the QSL time become arbitrarily small. Hence, the operator $\hat h$ can be approximated as $\hat h = i \exp(i\Delta \hat \sigma_x \tau_{\mathrm{QSL}}/2)\partial_\Delta\exp(-i\Delta \hat \sigma_x \tau_{\mathrm{QSL}}/2) + \mathcal{O}(\tau_{\mathrm{QSL}}^2) \sim \hat \sigma_x \tau_{\mathrm{QSL}}/2$. Therefore, the state maximizing the QFI is the spin-down state which happens to also be the ground state of the LZ model for $g\gg \Delta$. For that reason, optimal \emph{regular} quantum metrology becomes equivalent to our critical quantum metrology approach in the limit of small evolution times $T\ll 1/\Delta$ and $g\gg \Delta$.

The results above suggest that the critical slowing down near the avoided crossing prevents the QFI to reach or overcome the SQL. Therefore, we now consider a shortcut to adiabaticity, specifically CD driving, which allows to prepare a (critical) ground state in arbitrary short times. We add the CD term $\hat H_{\mathrm{CD}}$ to the LZ Hamiltonian ensuring effective adiabatic dynamics for any control field $g(t)$~\cite{berry2009transitionless}
\begin{align}
    \hat H_{\mathrm{CD}} = - \frac{\dot{g}(t)\tilde\Delta}{2[\tilde\Delta^2 +g(t)^2]}\hat \sigma_y,
\end{align}
where the dot notation $\dot g(t)$ indicates a time derivative. As mentioned previously the CD term involves the parameter $\Delta$ and is therefore not exactly realizable when the parameter is unknown. Therefore, the parameter $\Delta$ in the CD term is replaced by an initial estimate $\tilde \Delta$ which is then updated adaptively after each measurement~\cite{pang2017optimaltd}. The ramp function $g(t)$ can be chosen arbitrarily and therefore we set $g(t) = g_0 -(g_0-g_f)(t/T)^{1/5}$ which is fast when the energy gap is large and becomes slower close to the avoided crossing. In the following we focus on preparing the ground state at $g_f=\Delta$ for which the QFI is maximal and set $\tilde\Delta = \Delta=0.05$ which corresponds to the ideal case where the estimate of the unknown parameter coincides with the true value.

\begin{figure}[htb!]
  \centering
\includegraphics[width=1\textwidth]{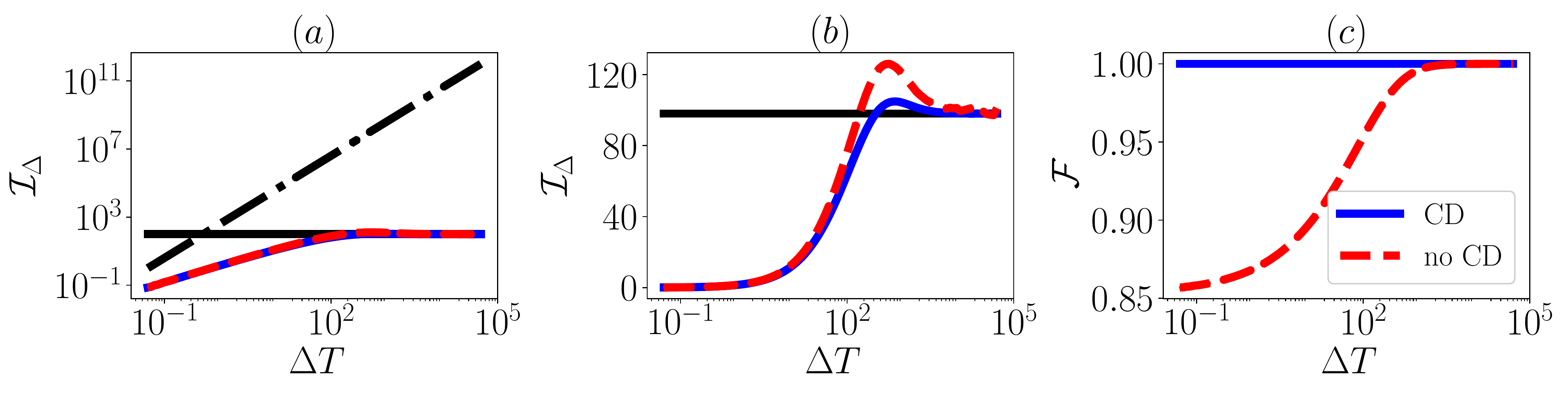}
\caption{The effect of CD driving on the QFI for the LZ model. The system is prepared in the spin-down state and then time evolved to the ground state at $g_f=\Delta=0.05$ close to the avoided crossing. Panels (a) and (b) show the QFI $\mathcal{I}_\Delta$ as a function of time duration $T$ of the drive expressed in the units of $\Delta$ in log-log scale (left) and log-linear scale (middle). The dashed red curve is the QFI for a sole ramp (without adding the CD term) while the solid blue line is the QFI for a ramp with CD driving. The dashed-dotted black line represents the HL. In the adiabatic limit, i.e.~for large evolution times $T$, the QFI attains the previously predicted value (solid black line). Panel (c) displays the fidelity $\mathcal{F}$ between the final time-evolved state and the target ground state for both cases of driving (with and without CD term) as a function of time.}
\label{fig:fig2}
\end{figure}

As shown in figure \ref{fig:fig2}(c) with the addition of the CD term the target ground state is indeed reached with unit fidelity in arbitrary short times (blue line) while driving without the CD term (dashed-red line) gives rise to excited states for short ramps away from the adiabatic limit. The QFI on the other hand now explicitly depends on time and is plotted in figure \ref{fig:fig2}(a),(b) for both cases of driving with and without the CD term. The critical quantum metrology scheme performs again considerably worse than the SQL (black dashed-dotted line) even though the critical ground state can be prepared in much shorter times [see figure \ref{fig:fig2}(a)]. Moreover, the achieved QFI quickly goes to zero for small evolution times and is able to surpass the previously calculated value of the QFI in the adiabatic limit (black solid line) only for intermediate times [see figure \ref{fig:fig2}(b)]. Therefore, \emph{regular} quantum metrology operated close to the SQL outperforms critical quantum metrology also in this case when using CD driving and hence when avoiding the problem of critical slowing down. Before discussing these results in Section~\ref{sec:dis}, we first analyze a more complex system that, unlike the LZ model, exhibits a quantum phase transition.
%
%
\subsection{Quantum Rabi model under the Schrieffer-Wolff transformation}
The quantum Rabi model is a finite-component system composed of a single two-level atom interacting with a single bosonic mode
\begin{align}
    \hat H_{\mathrm{QRM}} = \Delta \hat a^\dagger \hat a + \frac{\Omega}{2} \hat {\sigma_z} +\frac{g}{2}\left( \hat a^\dagger + \hat a \right) \hat \sigma_x.
\end{align}
where $\Delta$ is the frequency of the bosonic field represented by its creation and annihilation operators $\hat a^\dagger$ and $\hat a$, $\Omega$ is the energy splitting of a two-level atom represented by Pauli matrices $\hat \sigma_i$, and $g$ is the coupling parameter between the bosonic field and the two-level atom. In the limit of $\Delta/\Omega \rightarrow 0$, which can be considered as the thermodynamic limit~\cite{plenio2015rabimodelqpt}, the quantum Rabi model exhibits a superradiant phase transition at $g_c \equiv \sqrt{\Delta\Omega}$ that can be harnessed in critical quantum metrology~\cite{felicetti2020criticalqm}.

\begin{figure}[htb!]
  \centering
\includegraphics[width=1\textwidth]{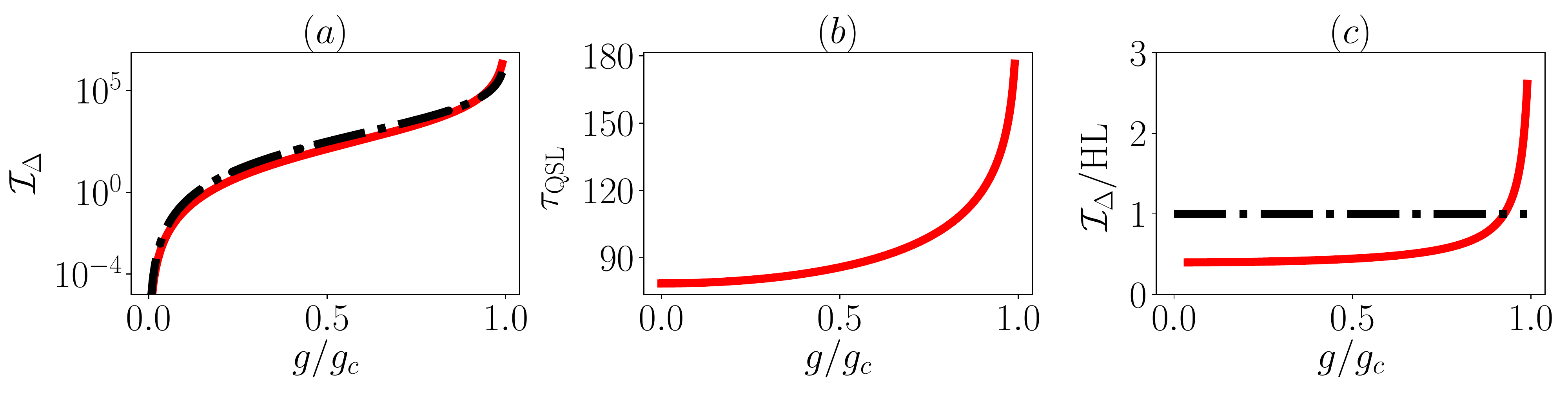}
\caption{QFI $\mathcal{I}_{\Delta}$ for the quantum Rabi model. Panel (a) presents the QFI as a function of $g/g_c$ for the case of critical parameter estimation (solid red line) and the HL for the case of \emph{regular} quantum metrology that could be potentially achieved if the protocol lasted for $\tau_{\mathrm{QSL}}(g)$ (black dashed-dotted line). Panel (b) depicts the QSL time $\tau_{\mathrm{QSL}}$. Panel (c) shows the QFI of critical quantum metrology (solid red line) normalized to the HL (black dashed-dotted line). In the simulations we set $\Delta = 0.01$ and $\Omega = 100$. The maximal value of $g$ shown in the plots is $0.99g_c$.}
\label{fig:fig3}
\end{figure}

To obtain an analytical expression for the QFI we need to compute the ground state of the system in an analytical form which is difficult for the general case~\cite{zhong2013analytical,you2015qrmgs}. However, in the suitable limit of $\Delta/\Omega \rightarrow 0$, the system can be diagonalized with the help of the Schrieffer-Wolff transformation $\hat U_{\mathrm{SW}} =\exp\{i (g/2\Omega) (\hat a^\dagger + \hat a)\hat \sigma_y\}$ which up to $\mathcal{O}(\Delta\sqrt{{\Delta}/{\Omega}})$ terms leads to
\begin{align}
    \hat U_{\mathrm{SW}}\hat H_{\mathrm{QRM}} \hat U_{\mathrm{SW}}^\dagger \equiv \hat H_{\mathrm{SW}} \simeq \Delta \hat a^\dagger \hat a + \frac{\Omega}{2} \hat \sigma_z +\frac{g^2}{4\Omega}\left( \hat a^\dagger + \hat a \right)^2 \hat \sigma_z.
\end{align}
This effective model (for clarity we will keep referring to it as the quantum Rabi model) can be diagonalized and the ground state is given by
\begin{align}
    |\psi_0\rangle = \hat S(\xi) |0 \rangle \otimes |\!\downarrow \,\rangle,
\end{align}
where $\hat S(\xi) \equiv \exp\{(\xi/2)(\hat a^\dagger)^2-(\xi^*/2)\hat a^2\}$ is the squeeze operator with $\xi = -\frac{1}{4} \ln\{1-(g/g_c)^2\}$ being the squeezing parameter which is real only for $g < g_c$. The latter condition restricts the validity of this ground state to the normal phase. Although an effective model for the superradiant phase can be derived~\cite{felicetti2020criticalqm}, in what follows we will focus on the normal phase only. Given the analytical expression for the ground state we can compute the QFI with respect to an unknown parameter which we assume to be $\Delta$. Near the critical point the QFI becomes~\cite{felicetti2020criticalqm}
\begin{align}\label{eq:QFIQRM}
    \mathcal{I}_{\Delta} \simeq  \frac{1}{32 \Delta^2(1-g/g_c)^2}.
\end{align}
The QFI diverges at the critical point $g = g_c$, where an arbitrarily large precision can be achieved [see red curve in figure~\ref{fig:fig3}(a)]. Similarly to the LZ case, the QFI above is attained when the (critical) ground state is prepared adiabatically. To include the required time resources for reaching a specific target ground state, we again use the QSL as a lower bound to the adiabatic evolution time. In the following we assume that the initial state is always given by the ground state of the Hamiltonian at $g=0$, i.e.~the vacuum state of the field and the spin-down state $\ket{0}\otimes\ket{\downarrow}$. The QSL time for achieving the ground state of the Hamiltonian for a different value of $g$ can be calculated according to the bang-off protocol (see Appendix \ref{appendix:C} for a detailed derivation) in which first a quantum kick is applied such that $g_{\mathrm{bang}}^2 = 2\Delta\Omega$ and $g_{\mathrm{bang}}^2 T_{\mathrm{bang}}/\Omega =- \ln(1-(g/g_c)^2)/2$, and subsequently one waits for a time $T_{\mathrm{off}} = \pi/(4\Delta)$ with $g_{\mathrm{off}}=0$, where $T_{\mathrm{bang}}$ and $T_{\mathrm{off}}$ is the duration of the bang and off time, respectively. The QSL is then given by $\tau_{\mathrm{QSL}} = T_{\mathrm{bang}} + T_{\mathrm{off}}$ and plotted in figure~\ref{fig:fig3}(b).

To compare the QFI in the critical quantum metrology scheme of equation \eqref{eq:QFIQRM} to the HL of \emph{regular} quantum metrology, we first need an expression for the latter, which requires some extra considerations. The quantum Rabi model does not conserve the number of photons and the ground state itself exhibits a different number of photons depending on the coupling parameter $g$. The HL can be computed for every $g$ by calculating the maximal sensitivity that can be achieved in single mode phase estimation~\cite{monras2006optimalphase} for a state with a fixed average number of photons $\langle n \rangle$, which turns out to be the squeezed vacuum state, i.e.~the ground state of the quantum Rabi model. The HL is then given by  $\mathcal{I}_\Delta = 8T^2(\langle n \rangle^2 +\langle n \rangle)$ where $T$ is the total evolution time. The HL after setting the time $T$ to the previously computed QSL time $\tau_{\mathrm{QSL}}(g)$ required for ground state preparation in critical quantum metrology is shown as a black dashed-dotted line in figure~\ref{fig:fig3}(a),(c). In contrast to the LZ model, if we could adiabatically evolve the quantum Rabi model within the QSL time, it would be possible to overcome the HL for states close to the critical point at $g\sim g_c$. We have to keep in mind though that the QSL time is a very optimistic lower bound on the actual required adiabatic evolution time. Hence, the achievable HL (black dashed-dotted line) when considering the true time resources of adiabatic state preparation will likely be larger than the QFI (red solid line) computed from equation \eqref{eq:QFIQRM}. However, these general considerations serve as a benchmark result and naively suggest that reducing the time of reaching the critical ground state might indeed lead to a potential advantage of critical quantum metrology for the quantum Rabi model.

Therefore, we next consider adding the appropriate CD term to the Hamiltonian (see Appendix \ref{appendix:B} for a derivation) given by
\begin{align}
\hat{H}_\mathrm{CD} = i \frac{g(t)\dot{g}(t)}{4\left(g_c^2 - g^2(t)\right)}\left[\left(\hat{a}^\dagger\right)^2 - \hat{a}^2\right].
\end{align}
In contrast to the LZ model, the CD term for the quantum Rabi model does not explicitly depend on the unknown parameter $\Delta$ (or an estimate $\tilde\Delta$ thereof). The dependence on the unknown parameter enters only through the expression for the critical coupling $\tilde{g}_c = \sqrt{\tilde\Delta\Omega}$ that we have to insert into the ramp function which we choose to be $g(t) = \sqrt{t/T}\tilde{g}_c$. We again consider the ideal case where the estimate matches the value of the unknown parameter, i.e.~$\tilde\Delta = \Delta = 0.01$.

\begin{figure}[htb!]
  \centering
\includegraphics[width=0.66\textwidth]{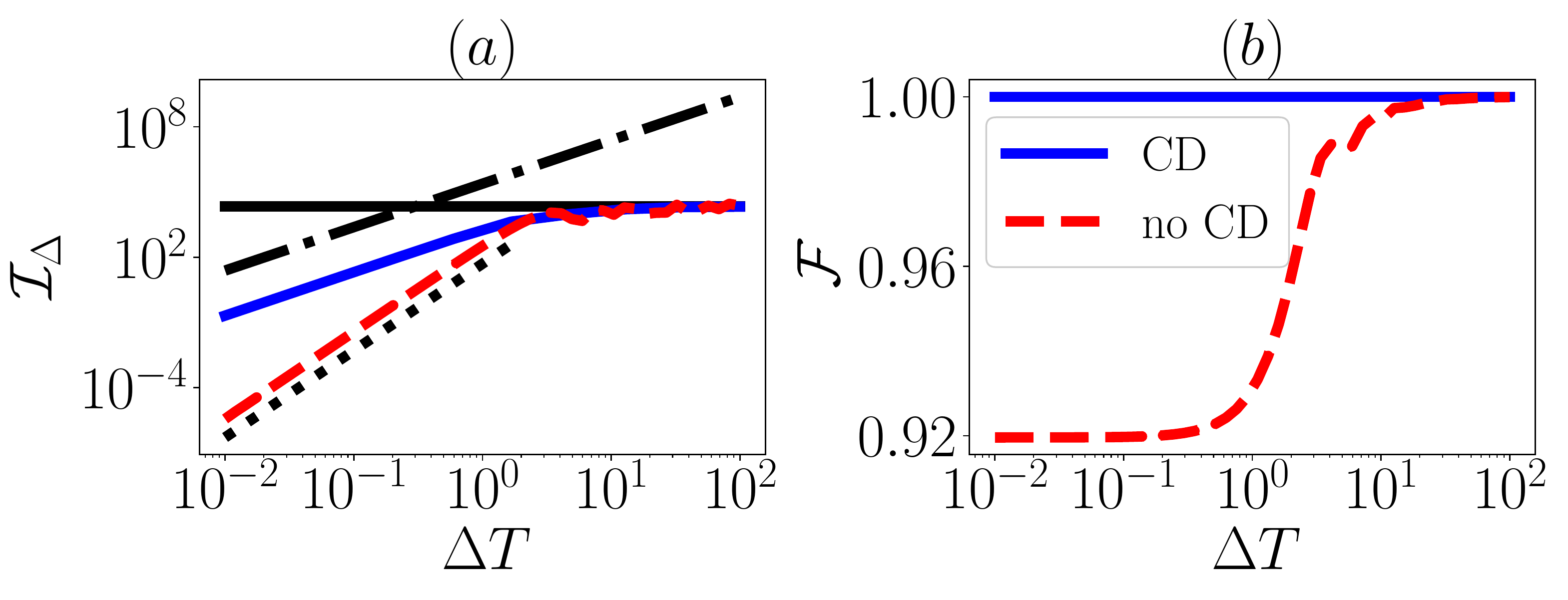}
\caption{The effect of CD driving on the QFI $\mathcal{I}_{\Delta}$ for the quantum Rabi model. In panel (a), the solid black line corresponds to the previously predicted adiabatic limit of the QFI for large driving times $T$, the dash-dotted black line is the HL as a function of the evolution time $T$, the dashed red line is the QFI for a sole ramp (without adding the CD term), and the solid blue line is the QFI for a ramp with CD driving. The black dotted line indicates a $T^4$ scaling of the QFI. Panel (b) shows the achieved fidelities $\mathcal{F}$ between the final time-evolved states and the target critical ground state as a function of time. In the simulations we set $\Delta = 0.01$, $\Omega = 100$, and $g_f/g_c = 0.9$.}
\label{fig:fig4}
\end{figure}

Figure~\ref{fig:fig4}(b) shows the achieved target ground state fidelities when driving the system with and without the CD term and indicates that critical ground state preparation can be reduced to arbitrary short times with the addition of the CD term as expected. However, the computed QFI for either of the driving protocols [plotted in figure~\ref{fig:fig4}(a)] is not able to overcome the HL (black dashed-dotted line) and reaches the adiabatic limit (solid black line) only for very large driving times. Together with the example of the LZ model these results confirm that shortcuts to adiabaticity cannot be used to saturate or beat the HL and therefore inevitably lead to lower sensitivities than performing optimal \emph{regular} quantum metrology using the same amount of time resources. 

%
%

\section{Discussion}
\label{sec:dis}
The arguments presented in Sections \ref{sec:qm} and \ref{sec:cqm} show that critical quantum metrology cannot yield better sensitivities than optimal \emph{regular} quantum metrology given the same time resources even when shortcuts to adiabaticity are applied which we illustrated using the two examples of the LZ and the quantum Rabi model. However, it does not yet explain why in these two cases critical quantum metrology using CD driving performs considerably worse away from the adiabatic limit, i.e.,~why the achievable QFI goes to zero for short evolution times even though the target ground state is always reached with almost unit fidelity. In the following we will provide an intuitive explanation for the poor performance of CD driving in quantum metrology.

\begin{figure}[htb!]
  \centering
\includegraphics[width=0.45\textwidth]{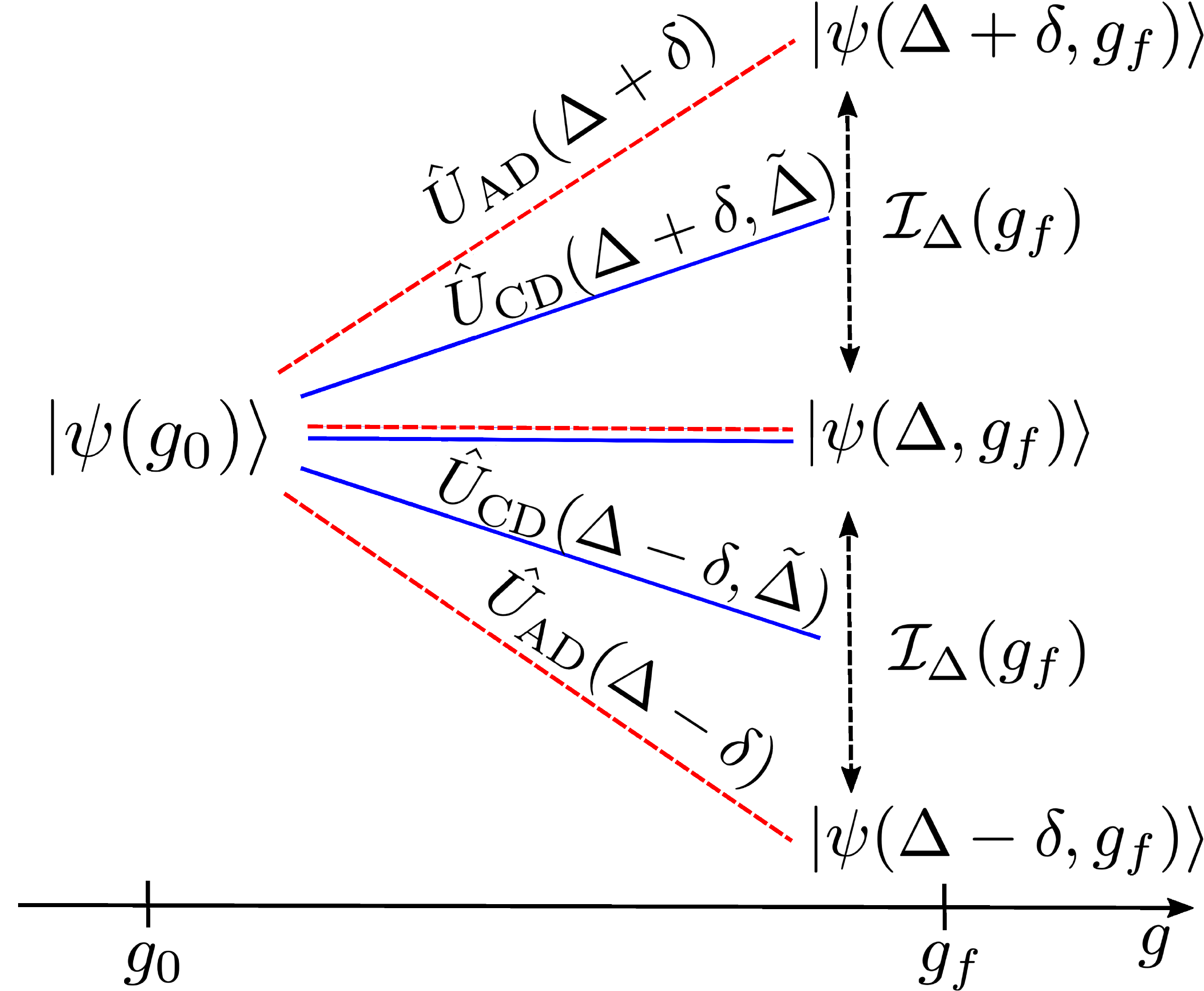}\hfill
\caption{The QFI $\mathcal{I}_{\Delta}$ can also be interpreted as a distance between quantum states in the parameter space of the unknown parameter $\Delta$~\cite{caves1994geomqs}. In the case of adiabatic state preparation (red-dashed lines), the final quantum states $\ket{\psi(\Delta, g_f)}$ and $\ket{\psi(\Delta+\delta, g_f)}$ reached after time evolution using the unknown parameter $\Delta$ and $\Delta+\delta$ respectively, are always the ground states of the corresponding Hamiltonians. Hence, the QFI can be computed through a consideration of the final (critical) Hamiltonian and its ground state alone. Yet, the time needed to adiabatically prepare the ground state at the critical point will in general diverge. Therefore, counter-diabatic driving is employed to reduce the time of reaching the final target ground state $\ket{\psi(\Delta, g_f)}$ (blue solid lines). However, the time evolved state $|\psi(\Delta+\delta, \tilde\Delta, g_f)\rangle$ when the unknown parameter is shifted by $\delta$ is no longer the ground state of the corresponding Hamiltonian, but rather a state moved closer towards the target ground state $\ket{\psi(\Delta, g_f)}$ for which the CD term was designed for. This gives rise to a larger overlap between those quantum states and in turn to a smaller QFI.}
\label{fig:fig5}
\end{figure}

The expression for the QFI $ \mathcal{I}_{\Delta} \equiv 4\left(\langle\partial_\Delta \psi_f |\partial_\Delta \psi_f\rangle - \langle \partial_\Delta \psi_f| \psi_f\rangle^2 \right)$ can be rewritten as
\begin{align}\label{eq:fidel}
	\mathcal{I}_{\Delta}  = \lim_{\delta \rightarrow 0} \frac{4}{\delta^2}\left(1-|\langle\psi_f(T, \Delta, \tilde \Delta, g_f)|\psi_f(T, \Delta+\delta, \tilde \Delta, g_f)\rangle|^2\right),
\end{align}
where $\delta$ is an infinitesimal change of the unknown parameter $\Delta$. Thus, the QFI reaches its maximum value when the overlap of the final states $|\psi_f(T, \Delta, \tilde\Delta)\rangle$ and $|\psi_f(T, \Delta+\delta, \tilde\Delta)\rangle$ is minimal which corresponds to a larger distance between these quantum states in the $\Delta$-parameter space and therefore an enhanced sensitivity (see figure \ref{fig:fig5} for an illustration). For adiabatic processes (hence large evolution times~$T$), the CD term becomes irrelevant, and both of the time-evolved states $|\psi_f(T, \Delta)\rangle$ and $|\psi_f(T, \Delta+\delta)\rangle$ in equation \eqref{eq:fidel} will be ground states of their respective final Hamiltonians. However, for short evolution times the CD term gains importance ensuring that $|\psi_f(T, \Delta, \tilde\Delta)\rangle$ is the ground state when $\tilde\Delta=\Delta$ is set appropriately. The state $|\psi_f(T, \Delta+\delta, \tilde\Delta)\rangle$ on the other hand is no longer the ground state of the Hamiltonian at $\Delta+\delta$. In fact, the CD term pushes the final state closer to the target ground state of the Hamiltonian at $\Delta=\tilde\Delta$ for which it has been designed for. This leads to an increase in the overlap of the two time-evolved states and in turn the QFI decreases illustrated by the blue line in figure \ref{fig:fig5}. Hence, for shorter evolution times $T$, the effect of the CD term will be larger, giving rise to smaller distances of the final time-evolved quantum states in the parameter space of the unknown parameter and therefore a lower sensitivity.

However, these results do not necessarily imply that shortcuts to adiabaticity are not useful in the context of quantum metrology or critical quantum metrology is impractical. Shortcuts to adiabaticity and related techniques can be employed to prepare suitable initial states~\cite{frerot2018qcm,hatomura2018shortcuts} or to lead a quantum state through a quantum path that maximizes the QFI~\cite{pang2017optimaltd}. In certain cases CD driving can still be beneficial and can give rise to a higher QFI than driving without any additional control (see figure \ref{fig:fig4}). On the other hand critical quantum metrology might be useful in cases where the optimal conditions for reaching the HL in the \emph{regular} quantum metrology setting cannot be easily achieved experimentally. Critical quantum metrology will also be advantageous if the initial state of the system of interest is already (close to) the critical ground state and therefore the associated time resources for preparing a critical state can be neglected which however, is not often the case in real experimental setups. The oft-cited super-Heisenberg scaling in critical quantum metrology achievable in critical quantum metrology $\mathcal{I}_{\Delta} \sim N^{>2}$~\cite{witkowska2020supersensitive} derives from the fact that the QFI is calculated without considering the adiabatic protocol time. When this time duration is included~\cite{Rams2018atthelimitsofcriticality,felicetti2020criticalqm}, the apparent super-Heisenberg scaling vanishes and the sensitivity is limited by the HL $\mathcal{I}_{\Delta} < N^2T^2$. 
%
%
\section{Conclusions}\label{sec:con}
In this work, we have shown that the HL achievable in \emph{regular} quantum metrology also poses an upper bound for the attainable QFI in critical quantum metrology when the preparation of the critical ground state is taken into account as well. However, as reaching the HL requires different states than instantaneous eigenstates of a critical system, optimal \emph{regular} quantum metrology is always superior to critical quantum metrology given the same amount of time resources. We confirmed that previous reports~\cite{felicetti2020criticalqm, witkowska2020supersensitive} on beating the HL in critical quantum metrology are a consequence of neglecting the time required to prepare a critical state~\cite{Rams2018atthelimitsofcriticality}. We have also shown that shortcuts to adiabaticity, specifically counter-diabatic driving, cannot be used to reach or overcome the HL, although they allow the critical ground state to be prepared in arbitrary short times. In fact, for the two examples considered here, CD driving in general leads to lower sensitivities than performing adiabatic quantum metrology without any extra control. 
%
%
\section{Acknowledgements}
The authors are pleased to acknowledge Thomas Busch, Thomás Fogarty, and Keerthy Menon for inspiring discussions. Simulations were performed using the open-source QuantumOptics.jl framework in Julia~\cite{kramer2018quantumoptics}. This work was supported by the Okinawa Institute of Science and Technology Graduate University. K.G. acknowledges support from the Japanese Society for the Promotion of Science (P19792).
%
\appendix


\section{Quantum Fisher information in critical quantum metrology}
\label{appendix:A}
In our critical quantum metrology scheme the system of interest is driven from the uncritical ground state of a Hamiltonian $\hat H(\Delta,g_i)$ to the critical ground state of a Hamiltonian $\hat H(\Delta,g_f)$, where $\Delta$ is the unknown parameter to be estimated, and $g_i,g_f$ are the initial and final values of the control field $g(t)$, respectively ($g_f$ is close to the critical point). The final state can therefore be calculated via the corresponding unitary evolution operator
\begin{align}
 \ket{\psi_f} \equiv \ket{\psi_f(\Delta,g_f)} = \hat U(T,\Delta,g_0,g_f) \ket{\psi_0(\Delta,g_0)},
\end{align}
with $\ket{\psi_0(\Delta,g_0)}$ being the unknown-parameter-dependent initial ground state and $T$ the total evolution time. Inserting the expression above into the definition of the QFI $ \mathcal{I}_{\Delta} \equiv 4\left(\langle\partial_\Delta \psi_f |\partial_\Delta \psi_f\rangle - \langle \partial_\Delta \psi_f| \psi_f\rangle^2 \right)$ and using $\hat h = i\hat U^\dagger\partial_\Delta\hat U $, yields
\begin{align}
\begin{split}
    \mathcal{I}_{\Delta} &= 4\left(\langle \psi_0 |\hat h^2| \psi_0\rangle - \langle  \psi_0|\hat h| \psi_0\rangle^2 \right) \\ 
    &\quad+ 4\left(\langle\partial_\Delta \psi_0 |\partial_\Delta \psi_0\rangle - \langle \partial_\Delta \psi_0| \psi_0\rangle^2 \right)\\  
    &\quad+ 4\left(\bracket{\psi_0}{\left(\partial_\Delta \hat U^\dagger\right)\hat U}{\partial_\Delta \psi_0} +\bracket{\partial_\Delta \psi_0}{\hat U^\dagger\left(\partial_\Delta \hat U\right)}{\psi_0}\right) \\ &\quad+2 \bracket{\psi_0}{\left(\partial_\Delta \hat U^\dagger\right)\hat U}{\psi_0}\braket{\partial_\Delta \psi_0}{\psi_0}\\ 
    &=\mathcal{I}_{\Delta}(\partial_\Delta \hat U) + \mathcal{I}_{\Delta}( \ket{\partial_\Delta \psi_0}) + \mathcal{I}_{\Delta}(\partial_\Delta \hat U, \ket{\partial_\Delta \psi_0}),
    \end{split}
\end{align}
If the initial ground state $\ket{\psi_0}$ is far from the critical state, its dependence on the parameter $\Delta$ is negligible, i.e.~$|\partial_\Delta \psi_0\rangle \simeq 0$, and the QFI becomes
\begin{align}
    \mathcal{I}_{\Delta} \simeq 4\left(\langle \psi_0 |\hat h^2| \psi_0\rangle - \langle  \psi_0|\hat h| \psi_0\rangle^2 \right).
\end{align}
Furthermore, if the time-evolution is adiabatic, that is, it follows the instantaneous ground state of the bare Hamiltonian $\hat H(\Delta,g(t))$, we obtain
\begin{align}
\begin{split}
    \mathcal{I}_{\Delta} &=4\left(\langle \psi_0 |\hat h^2| \psi_0\rangle - \langle  \psi_0|\hat h| \psi_0\rangle^2 \right) \\ &= 4\left(\langle\partial_\Delta \mathrm{GS}(\Delta,g_f) |\partial_\Delta \mathrm{GS}(\Delta,g_f)\rangle - \langle \partial_\Delta \mathrm{GS}(\Delta,g_f)| \mathrm{GS}(\Delta,g_f)\rangle^2 \right),
    \end{split}
\end{align}
where \ket{\mathrm{GS}(\Delta,g_f)} denotes the (critical) ground state of the Hamiltonian $\hat H(\Delta,g_f)$.

\section{Bang-off protocol for the critical ground state preparation of the quantum Rabi model under the Schrieffer-Wolff transformation}
\label{appendix:B}
The ground state of the quantum Rabi model is a squeezed vacuum state which is squeezed along the real axis of the Husimi Q function defined as~\cite{husimi1940some}
\begin{align}
    Q(\alpha) = \frac{1}{\pi}\langle \alpha | \hat \rho| \alpha \rangle,
\end{align}
where $\hat \rho$ is the density operator of the system which for pure states $|\psi\rangle$ becomes $\hat \rho = |\psi \rangle\! \langle \psi |$ and $|\alpha\rangle$ is a coherent state of the field. A squeezed vacuum state can be obtained from the quantum Rabi Hamiltonian by performing a proper pulse with the control parameter $g$. In order to make this explicit, let us rewrite the Hamiltonian 
\begin{align}
\begin{split}
\hat H &=\Delta \hat a^\dagger \hat a + \frac{\Omega}{2} \hat \sigma_z +\frac{g^2}{4\Omega}\left( \hat a^\dagger + \hat a \right)^2 \hat \sigma_z \\ 
  &= \Delta \hat a^\dagger \hat a + \frac{\Omega}{2} \hat \sigma_z +\frac{g^2}{4\Omega} \left( \left(\hat a^\dagger\right)^2 + \hat a^2 +2 \hat{a}^\dagger \hat{a} +  1 \right) \hat{\sigma}_z.
      \end{split}
\end{align}
Since the initial state is a spin-down state (an eigenstate of the $\hat \sigma_z$ operator), the $\hat \sigma_z$ operator can be replaced by $-1$, and the resultant constant terms can be dropped giving rise to
\begin{figure}[htb!]
  \centering
\includegraphics[width=1\textwidth]{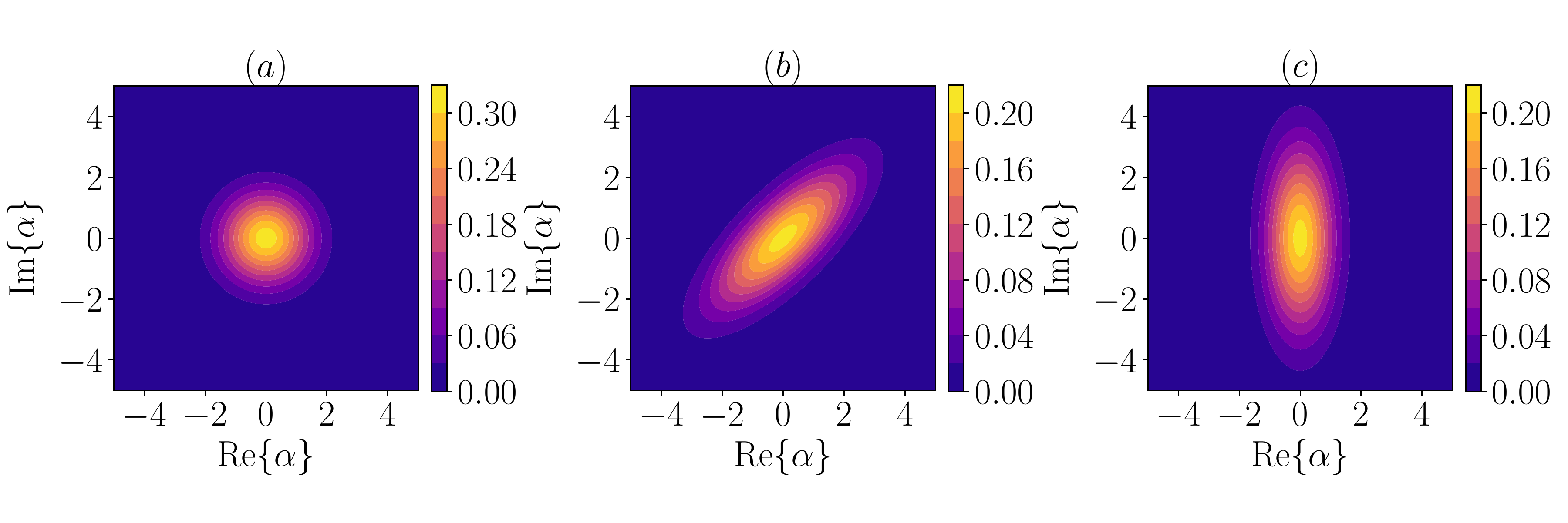}
\caption[fig6]{Husimi Q functions illustrating the bang-off protocol for critical ground state preparation. Panel (a) displays the initial vacuum state. Panel (b) shows the squeezed state after the bang step of the protocol which is not squeezed along the real axis and therefore does not yet correspond to the target ground state. Thus, during the off-step of the protocol a rotation of $\pi/4$ around the origin is applied yielding the critical ground state as shown in panel (c).}
\label{fig:fig6}
\end{figure}
\begin{align}
\hat H &= \left(\Delta- \frac{g^2}{2\Omega} \right) \hat a^\dagger \hat a  -\frac{g^2}{4\Omega} \left( \hat a^{\dagger 2} + \hat a^2 \right).
\end{align}
If we set $g = \sqrt{2\Delta \Omega}$, the first term vanishes, and we end up with a Hamiltonian
\begin{align}
\hat H &=   -\frac{\Delta}{2} \left( \hat a^{\dagger 2} + \hat a^2 \right),
\end{align}
which leads to a squeezing operator
 \begin{align}
 \hat U(t) = \exp\left( i t \frac{\Delta }{2}\left( \hat a^{\dagger2} + \hat a^2 \right) \right) = \exp\left( \frac{1}{2}\left(z^*\hat a^2 - z\hat a^{\dagger 2}  \right) \right) =\hat S(z),
 \end{align}
where $z = r e^{i \phi}$, with $r = t \Delta$ and $\phi = -\frac{\pi}{4}$ being the squeezing amplitude and squeezing direction, respectively. The ground state wavefunction of the quantum Rabi model is
\begin{align}
    |\psi_0\rangle = \exp\left(\frac{1}{2}\left(\xi^*\hat a^{\dagger2}-\xi\hat a^2\right)\right) |0 \rangle \otimes |\!\downarrow \,\rangle,
\end{align}
with $\xi = -\frac{1}{4} \ln\{1-(g/g_c)^2\}$. In order to prepare an equally squeezed state one has to set $t \Delta = -\frac{1}{4} \ln\{1-(g/g_c)^2\}$, and subsequently rotate the squeezed state such that it is squeezed along the real axis of the Husimi Q function. The rotation can be performed by turning off the control field such that the Hamiltonian becomes
\begin{align}
    \hat H = \Delta \hat a^\dagger \hat a,
\end{align}
for a time such that $t\Delta = \pi/4$. Therefore, the QSL time for this bang-off protocol is given by $\tau_{\mathrm{QSL}} = \left(\pi/4-\frac{1}{4} \ln\{1-(g/g_c)^2\} \right)/\Delta$. Note that $-\frac{1}{4} \ln\{1-(g/g_c)^2\} > 0 $. The sequence preparing the critical ground state is presented in figure~\ref{fig:fig6}. 
%
%
%

\section{Counter-diabatic driving for the quantum Rabi model under the Schrieffer-Wolff transformation}
\label{appendix:C}
The CD term for the shortcut to adiabaticity can be calculated as

\begin{align}
\hat{H}_\mathrm{CD} = i \sum_n\Big[ \ket{\dot n(t)}\!\bra{n(t)} - \braket{n(t)}{\dot n(t)}\ket{n(t)}\!\bra{n(t)}\Big],
\end{align}
where $\ket{n(t)}\equiv |\psi_n(t) \rangle$ is the instantaneous eigenstate of the bare Hamiltonian and $\ket{\dot n(t)}$ is its time derivative.
We have
\begin{align}
\ket{\dot n(t)} = \frac{g(t)\dot{g}(t)}{4\left(g_c^2 - g^2(t)\right)}\left[\left(\hat{a}^\dagger\right)^2 - \hat{a}^2\right]\ket{n(t)}
\end{align}
and therefore the overlap
\begin{align}
\begin{split}
\braket{n(t)}{\partial_t n(t)}\ &= \frac{g(t)\dot{g}(t)}{4\left(g_c^2 - g^2(t)\right)}\bracket{n}{\hat{S}^\dagger\left[\left(\hat{a}^\dagger\right)^2 - \hat{a}^2\right]\hat{S}}{n} \\
&= f(t)  \left(\bracket{n}{\hat{S}^\dagger\hat{a}^\dagger\hat{S}\hat{S}^\dagger\hat{a}^\dagger\hat{S}}{n} - \bracket{n}{\hat{S}^\dagger\hat{a}\hat{S}\hat{S}^\dagger\hat{a}\hat{S}}{n} \right)\\
&= f(t) \left[\bracket{n}{\left(\hat{a}^\dagger\cosh(r) + \hat{a}\sinh(r)\right)^2}{n} \right.
\\& \quad - \left. \bracket{n}{\left(\hat{a}\cosh(r) + \hat{a}^\dagger\sinh(r)\right)^2}{n}\right] \\
&= f(t)(2n + 1) \left[ \sinh(r)\cosh(r) - \sinh(r)\cosh(r)\right] = 0,
\end{split}
\end{align}
vanishes, where we have set $f(t) = g(t)\dot{g}(t)/\{[4(g_c^2-g^2(t)]\}$ and used 
\begin{align}
 \hat{S}^\dagger(\xi)\hat{a}\hat{S}(\xi) = \hat{a}\cosh(r) - \hat{a}^\dagger e^{i\vartheta}\sinh(r),
\end{align}
with $\xi = r e^{i\phi} = \frac{1}{4}|\ln(1-g^2(t)/g_c^2)|e^{i\pi}$ in our case. Using the completeness relation $\sum_n\ket{n(t)}\!\bra{n(t)}=\hat{\mathds{I}}$ of the instantaneous eigen-basis the CD term finally reads
\begin{align}
\hat{H}_\mathrm{CD} = i \frac{g(t)\dot{g}(t)}{4\left(g_c^2 - g^2(t)\right)}\left[\left(\hat{a}^\dagger\right)^2 - \hat{a}^2\right] \, .
\label{eq:counterdiabatic}
\end{align}


\end{document}